# Modeling Bike Share Station Activity: Effects of Nearby Businesses and Jobs on Trips to and from Stations

Xize Wang[a, *], Greg Lindsey[b], Jessica E. Schoner[c], and Andrew Harrison[d]

**Abstract**

The purpose of this research is to identify correlates of bike station activity for Nice Ride Minnesota, a bike share system in Minneapolis – St. Paul Metropolitan Area in Minnesota. We obtained the number of trips to and from each of the 116 bike share stations operating in 2011 from Nice Ride Minnesota. Data for independent variables included in models come from a variety of sources; including the 2010 US Census, the Metropolitan Council, a regional planning agency, and the cities of Minneapolis and St. Paul. We use log-linear and negative binomial regression models to evaluate the marginal effects of these factors on average daily station trips. Our models have high goodness of fit, and each of 13 independent variables is significant at the 10% level or higher. The number of trips at Nice Ride stations is associated with neighborhood socio demographics (i.e., age and race), proximity to the central business district, proximity to water, accessibility to trails, distance to other bike share stations, and measures of economic activity. Analysts can use these results to optimize bike share operations, locate new stations, and evaluate the potential of new bike share programs.

a. Price School of Public Policy, University of Southern California. Email: wangxize316@gmail.com. OCRID: 0000-0002-4861-6002.
b. Humphrey School of Public Affairs, University of Minnesota. Email: linds301@umn.edu.
c. Department of Civil Engineering, University of Minnesota. Email: jschoner@gmail.com.
d. San Francisco Municipal Transportation Agency. Email: andrew.harrison@sfmta.com.
* Corresponding author.

# INTRODUCTION

## Bike sharing

Bike share systems, which are designed to provide inexpensive bicycle rental at strategic locations throughout cities for quick, one-way trips from station to station, are among the world's fastest growing urban cycling innovations. As of 2010, there were an estimated 100 programs in approximately 125 cities worldwide with over 139,300 bicycles (Shaheen, Guzman, and Zhang 2010). New programs are being established annually (DeMaio 2009).

In the United States, bike sharing has also attained popularity, partially because of increasing concerns about environmental sustainability concerns and partially due to the success of prior programs (Shaheen, Guzman, and Zhang 2010). Through 2012, 15 bike-share systems claiming 172,070 users had been established in the US (Shaheen et al. 2012). Planners, policymakers, and advocates for cycling believe these systems provide many benefits to users: convenience, improved access to destinations and jobs, increased physical activity and health, increased mobility, positive environmental impacts, low cost, and more efficient use of infrastructure and other space (Shaheen, Guzman, and Zhang 2010). Bike share programs are also helpful to promote the positive image of cycling (Goodman, Green, and Woodcock 2014), since factors like social norms have a very high influence on attitudes regarding cycling (Underwood et al. 2014).

The history of bike share systems can be traced back to the free bike systems in 1993, through coin-deposit systems, and finally to the current IT-based systems. Despite operational improvements associated with these innovations, operational challenges remain, specifically with respect to system rebalancing (Shaheen, Guzman, and Zhang 2010). System rebalancing refers to the set of operational problems that occur because of unequal numbers of rides to and from stations. For example, as riders go from stations with low activity to popular destinations, docking stations at destinations can be filled, and bikes from areas of origin can be depleted. Managers need tools to predict which bike share stations will "gain" or "lose" bicycles. A new generation of the bike share programs – the demand responsive, multi-modal system – has been proposed to overcome such this limitation (Shaheen, Guzman, and Zhang 2010). However, these systems have not been deployed. Our research, which identifies factors associated with variation in bike-share station activity, provides insights essential to developing more efficient, user-friendly systems.

**Research needed to strengthen program operations**

Research on bike share systems is booming but most of the studies are grey literature reports focusing on descriptive analyses of bike share systems and their benefits, application rates, and user characteristics. Systematic, quantitative analyses of bike share operations or their impact on transportation systems remain limited (Fishman, Washington, and Haworth 2013). Published studies have shown that the bike share system users' main trip purposes are commuting and utility travel, including serving as the last-mile option for transit trips. As a result, bike-share systems may be helping to reduce the level of driving (Fishman, Washington, and Haworth 2013). Researchers also have documented the advantages of bike share as a transportation mode. In Lyon, France, for example, Jensen et al. (2010) measured the speed of the city's *Vélo'v* Bike share system using the 11.6 million bicycle trips in the city provided by its operator. The average speed of the trips of the shared bicycles during the rush hours was as high as 15 km/h, which is even higher than the average speed of vehicles in downtown European cities (10-15 km/h). This indicates the practical feasibility of reducing driving through bike share programs.

Looking forward, the focus of bike share system research will likely be on ongoing refinement of practical bike share system operations (e.g., predicting and meeting demand; improving redistribution of bicycles across stations) as well as on understanding customer behavior and local bike sharing impacts (Pucher, Buehler, and Seinen 2011). For example, researchers recently have reported studies of the optimization of the bike-sharing stations using mathematical network models (Lin and Yang 2011) and GIS-based models (García-Palomares, Gutiérrez, and Latorre 2012). Krykewycz et al. (2010) applied a two-stage model including a Geographic Information System (GIS) raster analysis and quantitative estimation to estimate the demand for bicycle-sharing system in peer European cities. A recent analysis of the August 2010 rentals of 65 Nice Ride stations in Minneapolis is most relevant for this study (Maurer 2012). Maurer (2012) found that the monthly rentals are related to trip generation factors (e.g. population size), trip attraction factors (e.g. number of "attracters" or destinations such as shopping centers or museums), and transportation network factors (e.g. existence of bikeways). Maurer (2012), however, only studied a narrow window (one month) of the operations of the Nice Ride (2014b) system, the system size since has doubled, and additional study is needed.

This paper is a part of a larger study to identify economic activity associated with the Nice Ride Minnesota bike share system in Minneapolis and St. Paul, Minnesota (Schoner et al. 2012). In this paper, we assess whether bike share station activity is associated with the presence of retail businesses and job accessibility, in addition to other socio demographic, built environment, and infrastructure variables. We test these relationships with different regression models.

**Correlates of bicycle traffic**

Insights into factors associated with use of bike share stations can be obtained from studies of general non-motorized traffic – bicycling and walking. For instance, many studies (e.g., (Hankey et al. 2012); (Lindsey et al. 2007)) have shown that non-motorized traffic volumes on particular segments of transportation networks are correlated with the socio demographic characteristics of nearby populations, characteristics of the built environment or urban form (Stinson and Bhat 2004, Coutts 2009), characteristics of linked transportation infrastructure (Maurer 2012) , as well as other factors, including weather (Wang et al. 2014). Research has shown that higher supply of bicycle and pedestrian infrastructure such as bike lanes, separate bike paths and "bicycle boulevards" helps to increase levels of bicycle and pedestrian travel (Dill 2009, Krizek, Barnes, and Thompson 2009). Higher population density also is related to higher levels of bicycling and walking (Handy et al. 2002). Another important factor is land use mix which has been shown to be associated with higher levels of active travel. Frank, Andresen, and Schmid (2004) found that people tended to bike and walk more when they lived at places with higher level of entropy, a measurement of the land use mix. Safety is also believed to be important to active travel levels. Higher crime rates in a neighborhood are usually related to lower active travel levels (Joh, Nguyen, and Boarnet 2012).

Because each of these general categories of variables– socio demographic, built environment, and transportation infrastructure – have been shown to be related to the non-motorized traffic volumes, it will be helpful for them to become control variables.

The next section of this paper is a brief introduction of the Nice Ride bike share system in Minneapolis-St. Paul, Minnesota. This section also summarizes our data and the methods we used to estimate our models. The following section presents the results, including the two models studies average total station activity for the 116 Nice Ride stations that were in operation in 2011. The section after discusses potential applications for our models in transportation planning and management. The final section summarizes our main findings, notes the limitations of our study, and suggests areas for future research.

**BIKE SHARE SYSTEM IN MINNEAPOLIS-ST. PAUL**

Since its beginning in spring 2010, the Nice Ride Minnesota bike share system has attracted large numbers of bike commuters, leisure cyclists, and new cyclists. The Nice Ride system initially had 65 stations in or near Minneapolis central business district (CBD), other commercial areas, and the University of Minnesota campus. By the end of the 2011 season, Nice Ride had expanded to include 116 stations throughout Minneapolis and St. Paul (Fig. 1), enrolled 3,693 subscribers, and provided bicycles for thousands of daily users. The total number of Nice Ride bike trips for 2011 surpassed 217,000.

[Fig. 1 Here]

Nice Ride Minnesota initially located bike share stations to maximize ridership through practical and intuitive understanding of factors believed to be associated with ridership. These factors included presence in the CBD, proximity to retail and commercial businesses, proximity to other destinations or features (university campus, libraries, lakes, parks, etc.), presence in higher density residential area, nearby bike infrastructure as well as other factors that have been shown to be associated with bicycling. These decisions generally were made with an understanding of neighborhood characteristics rather than detailed research on the presence of specific types of businesses. A community outreach program helped inform expansions into North Minneapolis area and the city of St. Paul. Specifically, Nice Ride was criticized for ignoring lower income, minority areas. To address equity concerns, Nice Ride then established eight additional stations outside the CBD in the city of Minneapolis. In addition, nine stations were located along a corridor in advance of construction of a new light rail line (i.e., the Central Corridor between Minneapolis and St. Paul), despite lower population densities and limited accessibility to stations.

To further optimize system performance, Nice Ride Minnesota has tracked use of stations, shared station trip records with a number of research organizations, and adjusted operations based on its records of station usage. As shown in Fig. 2, eight of the top 10 most used stations - measured in average trips per day - are located among high concentrations of retail land uses: six are near the Minneapolis CBD, one is near a major retail hub in a trendy Uptown neighborhood south of the CBD called Calhoun Square, and one is on the periphery of the University of Minnesota campus where many restaurants and shops are located. The remaining two are on the University of Minnesota campus. Conversely, the 10 least used stations are in areas with lower concentrations of retail and consumer-oriented businesses, including stations located in North Minneapolis for equity reasons and in the future light rail corridor.

[Fig. 2 Here]

Economic theories of transportation demand suggest that the availability of bicycle stations for inexpensive public use throughout a city should affect travel and consumption patterns, though the extent of these effects is not known. Bicycle share stations selectively increase accessibility to areas around each station by increasing the number of people who can reach particular places within reasonable travel times. Increases in accessibility also increase the potential for changes in local economic activity (i.e., consumption and spending). If bike share stations are associated with local economic activity, we hypothesize that the number of trips to and from stations should be positively associated with higher concentrations of nearby businesses, including food-related destinations, and accessibility to jobs.

To test our hypotheses quantitatively, we estimate log-linear and negative binomial models that use the total number of trips to and from stations as the dependent variables. Independent and control variables include indicators of economic activity (i.e., the number of food-related retail establishments and job accessibility), neighborhood socio-demographics, the built environment, transportation infrastructure, and other dummies specific to the study area.

**DATA AND METHODS**

We use regression modeling to identify the relationship between activity at 116 Nice Ride stations during the 2011 season and a set of 13 independent and control variables that reflect different relevant characteristics of the station area, which we define as area with a 0.4 km (¼ mile) network distance buffer for each station. Fig. 3 is an example of these station areas from the Lowry Hill East neighborhood in Minneapolis. The number of stations (116) is sufficiently large to reflect considerable variation in activity, and, given the number of independent variables (13), we have sufficient degrees of freedom for estimation.

[Fig. 3 Here]

**Correlates of nice ride station activity**

*Dependent variables*

For the purposes of this study, we define bike share "station activity" as the sum of trip origins and destinations. We will measure the average daily station activity in order to sort out the different operating dates among different stations. The descriptive statistics of the trips to and from the 116 Nice Ride stations are shown in Table 1. The "total trip" (both trip origins and trip destinations) numbers among the 116 stations range from 83 to 20,544 with an average of 3,749. The number of trip origins for each station range from the 37 to 9,843 with an average of 1,875. The number of trip destinations range from 39 as the lowest to 10,701 as the highest. The average number of trip destinations is 1,874. The stations with larger number of trip destinations also tend to have larger number of trip origins. The distributions of the total trips, trip origins and trip destinations are not distorted by different capacities among bike share stations, since the Nice Ride Minnesota system is keeping tracking the number of the remaining bikes in each location (via personal contact with Mr. Vars, from Nice Ride Minnesota). These statistics show that the distributions of the total trips are skewed by the stations with higher values, which, as discussed below, makes the logarithm transformation appropriate in the ordinary least square (OLS) modeling. Also, the fact that the trip variable is non-negative integer (i.e., a count) makes use of negative binomial models appropriate (Long and Freese 2005).

[Table 1 Here]

*Socio demographic variables*

Using the 2010 US Census data (US Census Bureau 2012), two socio demographic control variables are constructed for a 0.4 km (1/4 mile) network distance buffer around each station (Table 2). The variable *whitepct* controls for the racial structure of the analytical units, specifically, the proportion of the residents who are white/Caucasian. The variable *ynoldpct* controls for the age structure of the analytical units; it is the proportion of the residents who are younger than five or older than 64. We expect station activity positively correlates with a larger percentages of Caucasian and middle-aged people.

[Table 2 Here]

*Built environment variables*

To control for the effects of the built environment around each station on station activity, we constructed variables for proximity to various notable destinations that are believed to be trip attractors or generators (Table 2). The variables *diswater*, *discbd*, and *dispark* measure the proximity of the station in kilometers to the nearest lake or river, downtown Minneapolis or St. Paul CBD, and park. The *campus* dummy variable indicates whether the station is located on the University of Minnesota campus. We expect station activity to be positively associated with being closer proximity to parks, water, and downtown, and campus locations.

*Transportation infrastructure variables*

We hypothesize that transportation infrastructure such as bike trails will support the Nice Ride station activities since it increases access to individual stations (Table 2). The variable *trail* indicates whether a paved trail exists in the 0.4 km (1/4 mile) station network buffer area. There is also a base case which indicates no paved trail existence within the station area. The variable *neardis* measures the distance to the next nearest Nice Ride station. Variable *opendate*, which measures the operating dates of the stations of 2011, serves as the exposure variable to ensure the variation among the Nice Ride station activity is not because some stations were not open for the whole season. We expect station activity to be positively related to the presence of paved trails and the proximity to the next nearest bike share station.

*Economic activity variables*

After controlling for the socio demographic, built environment and transportation infrastructure variables described previously, we isolate the marginal relationship between indicators of economic activity and station activity (trips). The variable *access* calculates job accessibility, measured as the total number of jobs accessible within a 30 minute transit ride (i.e., bus or light rail) from the station using 2006 transit and employment data as per (Fan, Guthrie, and Levinson 2012). We use jobs accessible via transit because we are particularly interested in the last mile problem for transit users. We chose an accessibility measure that incorporates both transit and jobs rather than a variable such as number of bus stops because we hypothesize that the accessibility measure will reflect the relationship between bike share and transit better. The variable *food* indicates the total number of businesses categorized "food" within a 0.2km (1/8 mile) walking distance buffer around the station. These businesses were identified using the data from the Census Bureau and categorized by NAICS according to the protocol developed by Horning, El-Geneiday, and Krizek (2008) for measuring non-motorized accessibility. Both of the two economic activity variables are expected to be positively associated with 2011 station activity. We also tested also tested a measure defined as total businesses in the station area but this variable was not associated with station use and is not reported here.

*Controlling factors*

Three controlling factors, in addition to the variables discussed above, are included to better estimate the effects (Table 2). The variable *northmpls* is a dummy variable indicating whether the stations are located at the North Minneapolis area. These stations were installed for reasons of equity, rather than network optimization considerations, or because Nice Ride managers anticipated high levels of use. The variable *cclrt* is a dummy variable indicating whether the stations are located at areas in the future Central Light Rail Transit Corridor but were heavily affected by the ongoing construction of the project. Among other factors, the population and destination densities in the corridor are different than in and near the Minneapolis CBD. The variable *open2010* is a dummy variable that indicates the stations first opened in 2010 rather than 2011. The station activity is expected to be negatively associated with the variables *northmpls* and *cclrt*, but to be positively associated with the variable *open2010*. The variable *opendate* indicates the number of days operated during 2011, which is the exposure variable of the negative binomial regression.

**Model development and estimation**

We use both the log-linear OLS regression and negative binomial regression to estimate the station activity models. We use the logarithm form to estimate the average station activity since its distribution is skewed to the right-hand side. We use negative binomial regression to estimate the total station activity counts since they are non-negative integer values.

Similar to Poisson distribution, the dependent variable *y* in negative binomial regression model is also a non-negative integer. The probability when y equals *m* conditioning on the linear combination of $x_1$, $x_2$… and a parameter $\lambda$ is as formula (1) below (Long and Freese 2005):

$$P(y = m | x_1, x_2, \dots) = \frac{e^{-\lambda} \times \lambda^m}{m!} \qquad (1)$$

The negative binomial regression model assumes the mean of $y$ is $\lambda$ and the variance of $y$ is $\lambda + \alpha\lambda^2$. Maximum Likelihood Estimation (MLE) method is used to estimate $\alpha$ and the $\beta$s of the following generalized linear model as formula (2):

$$\ln \lambda = \beta_0 + \beta_1 \times x_1 + \beta_2 \times x_2 + \cdots \qquad (2)$$

If the observations of dependent variable $y$ are from time lengths, we need to add an exposure factor to control the effect of the different time periods (Long and Freese 2005).

Because the dependent variables in both of the two models are transformed into the natural logarithm form, the marginal effect of the independent variables will be different from that in basic OLS models. Specifically, if the coefficient of an independent variable is β, an increase of the variable by one unit is correlated with 100*(exp(β)-1)% increase of the dependent variables, or to its exp(β) times. The adjusted R-square is a measurement of the goodness-of-fit of the OLS model, and the Cox-Snell Pseudo R-square is a measurement of the goodness-of-fit of the negative binomial model.

The two models, estimated by Stata and SPost 9 (Long and Freese 2005), are:

- Log-linear model: The dependent variable is the natural logarithm of total station activity (trip origins plus trip destinations) per operating day of the 2011 season.
- Negative binomial model: The dependent variable is the total station activity (trip origins plus trip destinations) of the 2011 season, with an exposure time controlling variable *opendate*.

In both of the models, the effects on dependent variables measured are the average daily trips, rather than total daily trips. We also estimated models for average daily origins and departures; we do not report them here because the results were essentially the same as the models for average daily trips and because average daily trips better reflects overall station activity.

**RESULTS**

The results of the two models are shown in Table 3. Model 1 has a very high fit: the adjusted R-square value is 0.847, indicating nearly 85% of the variation in trip activity across stations can be explained by these variables. The F-statistic of Model 1 (80.14) is significant at a 0.0001 level, indicating the joint significance of the variables and the reliability of the adjusted R-square value. The pseudo R-square for Model 2 is 0.863, while it is not directly comparable to the adjusted R-square of the log-linear model. However, the fact that it has an even higher significance than the log-linear model indicates that it also has a very good fit. The likelihood ratio Chi-square value of Model 2 (230.27) is also significant at a 0.0001 level, it also supports the good fit of the model. In addition, the significance of the dispersion factor confirms that the negative binomial regression model is theoretically sounder than a Poisson model.

[Table 3 Here]

Overall, in both models, all the independent variables are significant at the 10% significance level, and the majority are significant at the 5% or 1% level. The fact that all independent variables are significant underscores the strength of our model fit and is indication of the validity of the underlying, hypothesized theoretical relationships. The signs of all but one coefficient are in the expected direction. The marginal effects of the two models are quite similar, indicating very robust estimations. The results of the tests of joint significance (F test and likelihood ratio test) confirm the reliability of these results.

Both of the socio demographic variables are significant at a 10% significance level at the least. The coefficient of *whitepct* is positive in both models, indicating that a higher activity level of a bike share station is positively related to a higher share of Caucasians in the population in the station area. Specifically, a 1% increase of the share of the white population at the 0.4 km (¼ mile) station buffer area is related to a 1.4-1.5% increase of the daily station activity. The coefficient of *ynoldpct* is negative, showing that the share of middle-aged residents is related to a higher level of station activity. Specifically, a 1% increase of the share of the population older than 64 or less than 5 in the buffer zone is associated with a 1.0-1.4% decrease of the bike-share station activities.

All of the four built-environment variables are significant at least at a 10% significance level. As hypothesized, there is a negative effect of *diswater*, *discbd* and *dispark* on the 2011 daily average Nice Ride station activities: bike share stations nearer to water bodies, to CBDs and to the parks have higher levels of activities. That implicitly shows people tend to use Nice Ride bikes for recreational purposes more frequently around rivers, lakes and parks; while they tend to use those shared bikes for commuting purposes in downtown areas. For instance, being one kilometer nearer to the CBD of Minneapolis or St. Paul is associated with an 11.5 – 11.6% increase of the bike share station use. Plus, at the University of Minnesota campuses, the average daily station trips are 42.6% or 48.4% higher, according to the two models.

The transportation infrastructure variables also play a significant role in explaining bike share station activity levels. *Trail* has a positive effect, indicating that 50.5% or 46.7% more trips occurred at stations connecting to paved trails. There is a positive effect of *neardis*, indicating that the stations within 1 kilometer of other stations tend to have on average 90.2% or 95.0% less trips. The sign of the coefficient, however, is not consistent with the hypothesis. This result might be because stations in close proximity may serve the same group of people and reduce the usage of individual stations.

As to the economic activity variables of interest, both are significant at a 5% level. The stations in areas with 1,000 more jobs connected via transit (variable *access*) tend to have 0.8% or 0.9% more bike share trips. The variable *food* is also significant, showing that the number of bike share trips at a station is strongly and positively correlated with the number of restaurants, cafeterias, and cafes around them. Specifically, the presence of one additional food related business is correlated with 1.7% more station trips. One interesting finding is that the marginal effects of the two models are consistent. As noted previously, we also tested the effect of total businesses in the station area but found no significant effects and so did not include results.

Finally, the three control factors are all significant at 5% significance level. We found the stations in North Minneapolis, which were installed for equity concerns average 37.6% or 42.0% lower level of station activity compared to other stations, according to the log-linear model and the negative binomial models, respectively. The stations heavily affected by the Central Corridor Light Rail Transit project have 38.2% or 40.7% lower station activity levels compared to other stations, according to the two models. The stations first opened in 2010 have 61.8% or 65.7% higher levels of station activities compared to the stations opened in 2011. In sum, the models are able to capture the wide variation in station activity associated with the stations even though different criteria were used for siting them.

**OBSERVATIONS, APPLICATIONS, AND CONCLUSIONS**

Our results provide new insights into factors that correlate with bike share station activity and can be used by managers and planners to optimize systems, locate new stations, and explore the feasibility of new systems. Bike share stations are unique within transportation systems, designed to provide people efficient, new options for multiple short rides. Bike share programs theoretically are best suited for locations with higher population densities and higher numbers of destinations that can be easily accessed. Nice Ride managers have not limited the location of stations to these types of areas, and in fact have stations for reasons of equity even though high activity levels were not anticipated. The results affirm initial decisions by Nice Ride managers to focus on locating stations near the Minneapolis CBD, on campuses, near parks and water bodies, and with access to off-street paths. These variables were highly significant in both models and associated with the largest marginal effects on station use. They also indicate that Nice Ride serves both utilitarian and recreational purposes. Socio demographic characteristics (e.g., age, race) of station areas are associated with station use, as are economic variables (e.g., access to jobs and proximity of food establishments) but their marginal effects on station use may be smaller depending on the variation across station areas. For example, the marginal effect of the presence of a paved urban trail in the station area on station use is approximately 50%, while the presence of an additional food establishment in the station area increases station use by about 1.7%. As noted, stations located specifically for reasons of equity have, as anticipated, significantly fewer trips originating or arriving at them.

The unanticipated negative coefficient on the measure of proximity to other bike share stations indicates that over-saturation of stations may present a challenge for system management. Relocation, monitoring of station use, and re-estimation of models may provide insights into the optimal number of stations, which can make the overall system more efficient.

While Nice Ride managers already know in real-time how use varies across stations, our results provide new information about why station use varies. The models also can be used to predict the effects of exogenous changes (e.g., the effects on campus stations when universities are not in session), creation of a cycle track, or the opening of new food-related businesses. If more detailed information is needed to improve logistics and system balancing, models can be estimated for origins and destinations separately.

To assist with siting of new stations or opening new systems, planners can predict station use at particular locations by obtaining values for the various variables and estimating the equations. System-wide effects of adding new stations or removing exiting ones can be modeled to help optimize the number of stations and determine priorities or sequences for new stations. Insights from use of these models also may inform marketing campaigns to increase use or outreach to potential sponsors of stations.

In sum, our study fills a gap in research on bike share systems by providing an empirical modeling framework to evaluate the correlates of bike share station activity using data from a comprehensive time window. Our models provide guidance for both practitioners and researchers interested in bike share system management and can inform transportation policy more generally. The fact that our model includes sites established for reasons of equity independent of the goal of maximizing use means that managers and policy makers can assess the consequences of siting stations in a variety of different locations.

One limitation of our research is that our models were estimated with data from only one system (i.e., Nice Ride). It would be interesting and useful to validate our model by testing it with data from other systems. Integration of data across systems would also help to identify new factors associated with bike share use.

While our research provides new insights into bike share station use and system operations, it also raises new questions and identifies needs for additional research. One limitation of our models is that because they involve cross-sectional analyses, we are able to show only association and not causation. Use of models in a controlled experiment (e.g., with a pre-post design) would yield additional useful insights (e.g., see Funderburg et al. (2010)). Additional analyses of the sensitivity of the models to size of station area and to different measures of proximity to other stations also would be interesting. For example, use of gravity models could be integrated into the modeling approach. Finally, additional research into the nature of specific types of retail and food establishments that affect use could provide important information for bike share managers.

## ACKNOWLEDGEMENTS

We greatly appreciate the financial support from the Bikes Belong Foundation. We also thank Mr. Mitch Vars from Nice Ride Minnesota, Mr. Tony Hull from Transit for Livable Communities and Ms. Mary Ann Murphy at University of Southern California for their help and support.

Schoner, Jessica E., Andrew Harrison, Xize Wang, and Greg Lindsey. 2012. Sharing to Grow: Economic Activity Associated with Nice Ride Bike Share Stations. Denver, CO: Bikes Belong Foundation.
Shaheen, Susan A, Stacey Guzman, and Hua Zhang. 2010. "Bikesharing in Europe, the Americas, and Asia." *Transportation Research Record: Journal of the Transportation Research Board* 2143 (1):159-167.
Shaheen, Susan A, E Martin, A Cohen, and R Finson. 2012. Public bikesharing in North America: Early operator and user understanding. Mineta Transportation Institute San Jose, CA.
Stinson, Monique, and Chandra Bhat. 2004. "Frequency of Bicycle Commuting: Internet-Based Survey Analysis." *Transportation Research Record: Journal of the Transportation Research Board* 1878 (-1):122-130. doi: 10.3141/1878-15.
Underwood, Sarah K, Susan L Handy, Debora A Paterniti, and Amy E Lee. 2014. "Why do teens abandon bicycling? A retrospective look at attitudes and behaviors." *Journal of Transport & Health*.
US Census Bureau. 2012. 2010 Decennial Census.
Wang, Xize, Greg Lindsey, Steve Hankey, and Kris Hoff. 2014. "Estimating mixed-mode urban trail traffic using negative binomial regression models." *Journal of Urban Planning and Development* 140 (1).
21

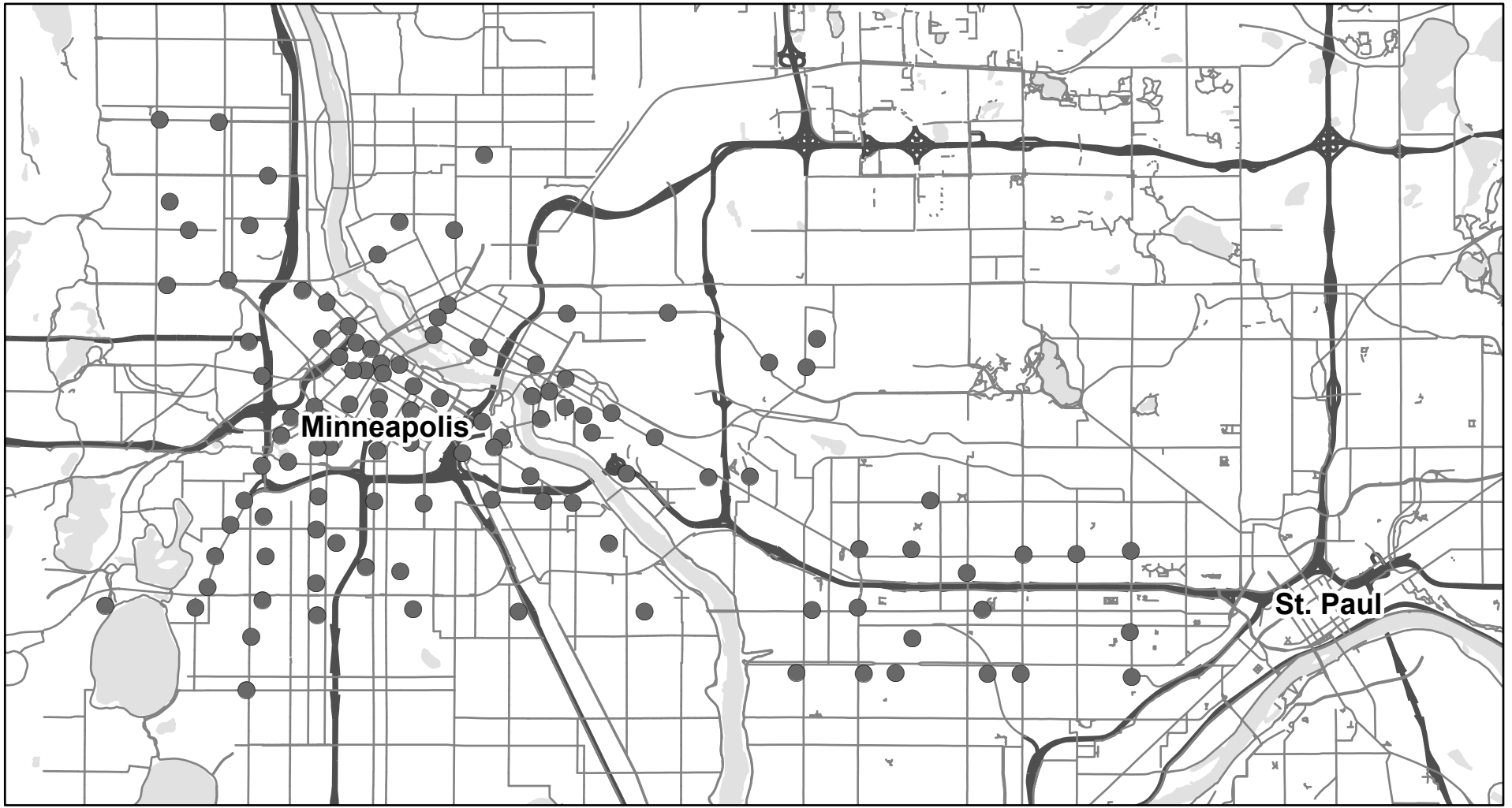

**Fig. 1** - Nice Ride Stations by the End of the 2011 Season (data from Nice Ride Minnesota 2014; MetroGIS 2014)

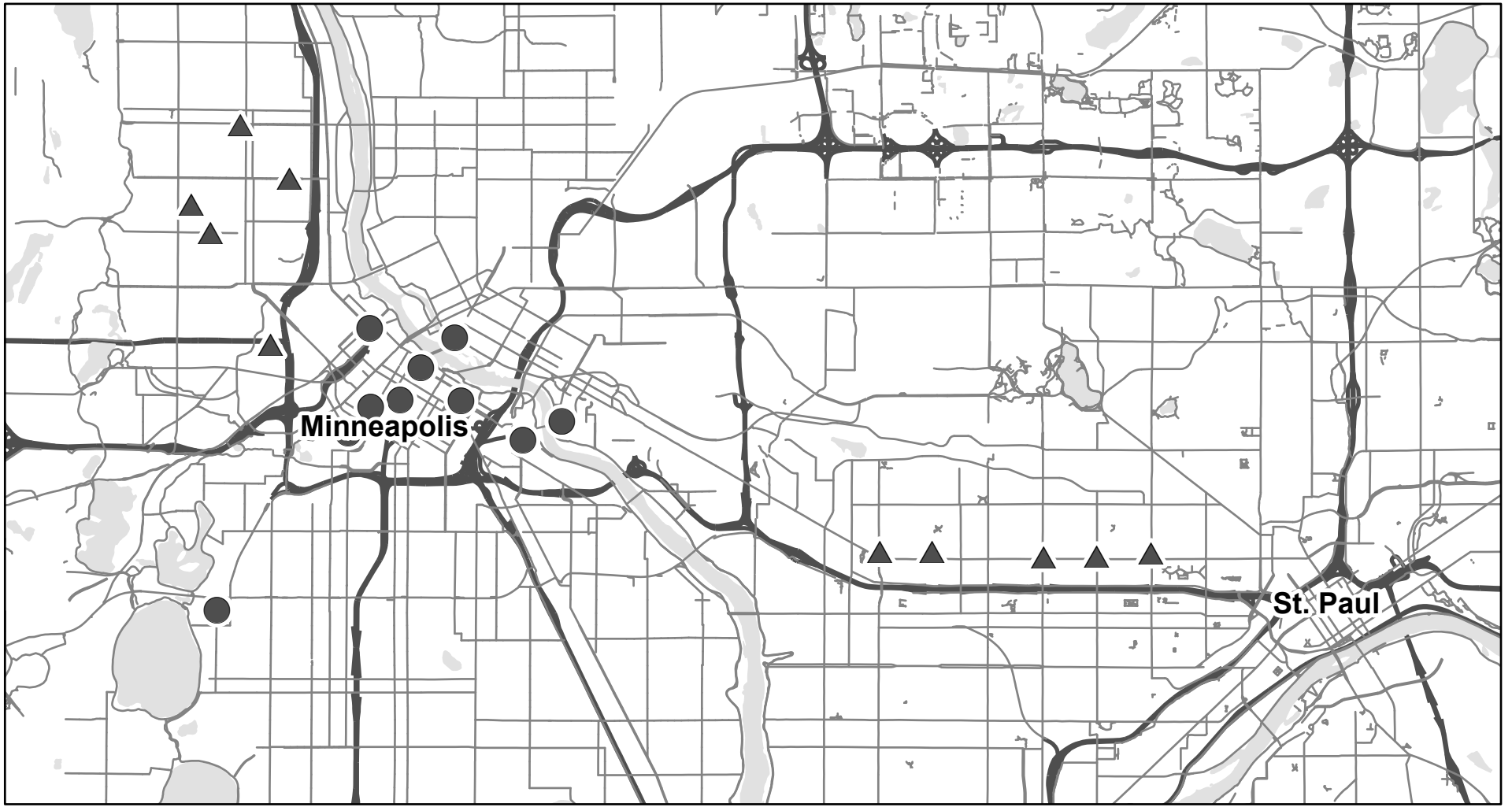

**Fig. 2** - Top 10 Origin and Destination Stations in 2011 (data from Nice Ride Minnesota 2014; MetroGIS 2014)

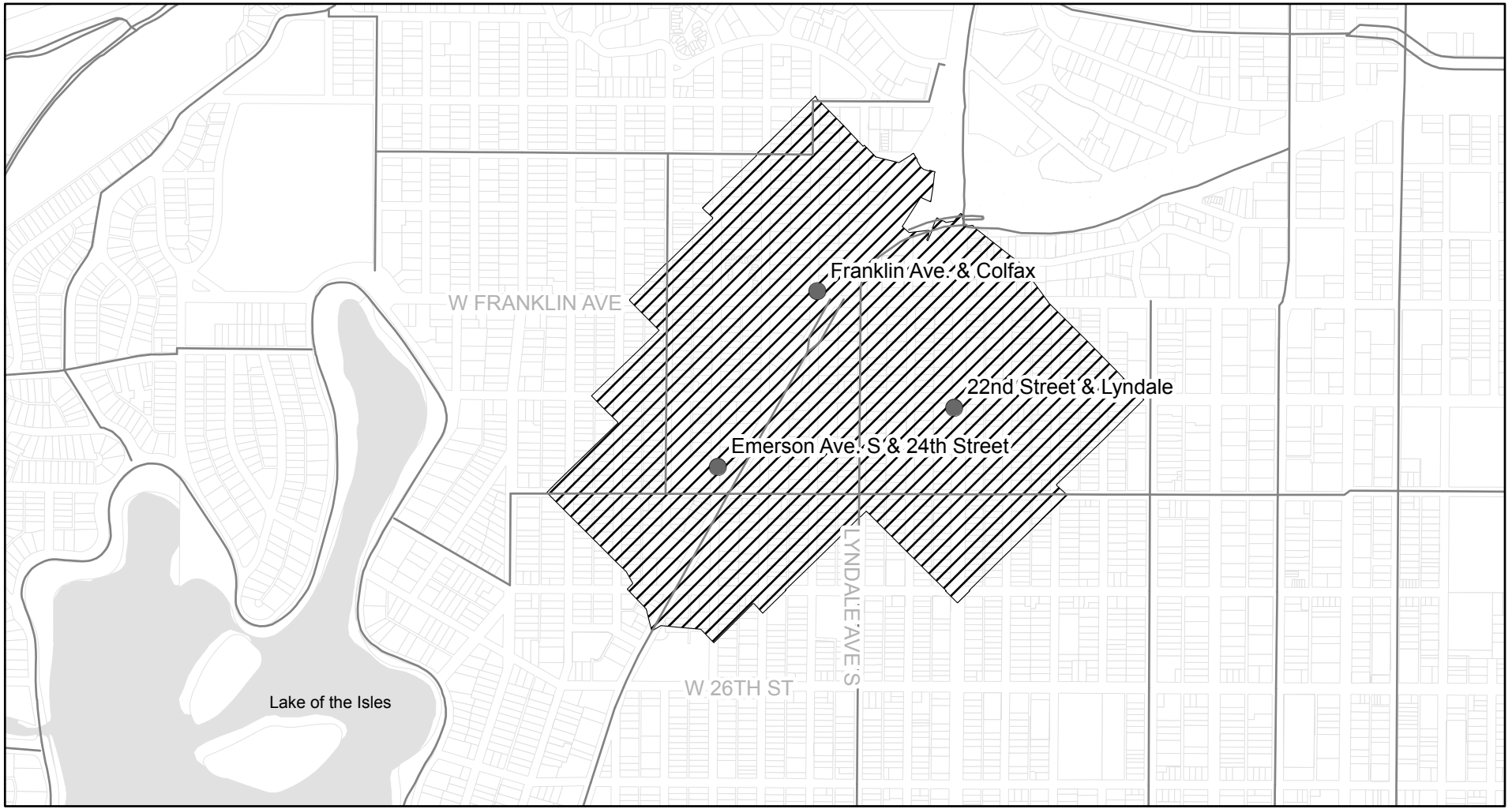

**Fig. 3** - 400-Meter Walking Distance Buffer around Lowry Hill East Stations
(data from Nice Ride Minnesota 2014; MetroGIS 2014)

**Table 1. Descriptive Statistics of Nice Ride Station Activity (for the year 2011)**

| Trip type | Average | Maximum | Minimum |
|---|---|---|---|
| Trips per day | 19.47 | 96.45 | 0.916 |
| Total trips | 3,749 | 20,544 | 83 |
| Trip origins | 1,875 | 9,843 | 37 |
| Trip destinations | 1,874 | 10,701 | 39 |

Table 2. Independent Variables Used in Modeling the Usage Frequency of the Nice Ride Stations.

| variable | Description | mean | Units/notes | exp. sign |
|---|---|---|---|---|
| | *Social Demographic Variables* | | | |
| whitepct | Percentage of the residents which are white/Caucasian. [1] | 60.51 | Unit: all census blocks interacted with the 0.4 km (¼ mile) station buffer area. | + |
| ynoldpct | Percentage of the residents which are less than 5 or more than 64 years old.[1] | 13.04 | Unit: all census blocks interacting with the 0.4 km (¼ mile) station buffer area. | - |
| | *Built Environment Variables* | | | |
| diswater | Distance to the nearest lake or river. [2] | 0.9973 | Unit: km. | - |
| discbd | Distance to the nearest CBD of each city for the bike share stations in Minneapolis or St Pau. [2] | 2.904 | Unit: km. CBD is defined to the centroid of the downtown reduced transit fee areas defined by Metro Transit. | - |
| dispark | Distance to the nearest park land use type.[2] | 0.2366 | Unit: km. Park is defined as land use type 170. | - |
| campus | Station at the University of Minnesota Campus. [2] | 0.1293 | Equals 1 if the station is on the University of Minnesota Minneapolis or St. Paul campuses, else 0. | + |
| | *Transportation Infrastructure Variables* | | | |
| trail | Presence of the paved trail in the station area. [2] | 0.2672 | Equals 1 if paved trail exists inside the 0.4km (¼ mile) station area buffer, else 0. | + |
| neardis | Distance to the nearest station. | 0.5142 | Unit: km. | - |
| opdate | Days of operation of the station in 2011. | 167.9 | The last day of service in 2011 is on Nov. 6th, 2011. | + |
| | *Economic Activity Variables* | | | |
| access | Total jobs in 30 minutes' transit accessibility in 2006.[3] | 37.24 | Unit: 1,000 jobs, includes all census blocks interacted with the 0.4km (¼ mile) station buffer area. | + |
| food | Total number of businesses in "food" category.[4] | 7.545 | Unit: 0.2 km (1/8 mile) station buffer area. | + |
| | *Controlling Factors* | | | |
| northmpls | The stations in the North Minneapolis established mainly on spatial equality criteria.[5] | 0.06897 | Equals 1 if located in the North Minneapolis area, equals 0 if not. | - |
| open2010 | Dummy variables for the stations opened in 2010. | 0.5603 | Equals 1 if the station was first opened in 2010, 0 if not. | + |
| cclrt | The stations heavily affected by the ongoing LRT construction.[5] | 0.07759 | Equals 1 if located in the areas affected by the Central Corridor construction, 0 else. | - |

[1] Data source: 2010 Census (US Census Bureau, 2012). [2] GIS data from the Metropolitan Council. [3] Data from Fan et al. 2012 [4] Data from US Census Bureau [5] selected by the authors' assessments.

### Table 3. Station Activity Regression Models

| variables | Model 1 (log-linear model) | | | Model 2 (Negative Binomial Model) | | |
|---|---|---|---|---|---|---|
| | coefficient | std. error | marginal | coefficient | std. error | marginal |
| **Dependent variable** | | | | | | |
| | Station activity (origin + destination) per opening day | | | Station activity (origin + destination) per opening day | | |
| **Social demographic** | | | | | | |
| whitepct | 0.015*** | [0.003] | 1.5% | 0.014*** | [0.002] | 1.4% |
| ynoldpct | -0.014* | [0.007] | -1.4% | -0.010* | [0.006] | -1.0% |
| **Build Environment** | | | | | | |
| diswater | -0.421*** | [0.086] | -34.4% | -0.444*** | [0.088] | -35.9% |
| discbd | -0.123*** | [0.043] | -11.6% | -0.122*** | [0.042] | -11.5% |
| dispark | -0.486* | [0.270] | -38.5% | -0.485* | [0.253] | -38.4% |
| campus | 0.355* | [0.193] | 42.6% | 0.395*** | [0.153] | 48.4% |
| **Transportation Infrastructure** | | | | | | |
| Trail | 0.409*** | [0.120] | 50.5% | 0.383*** | [0.099] | 46.7% |
| neardis | 0.668*** | [0.228] | 95.0% | 0.643*** | [0.227] | 90.2% |
| **Economic Activity** | | | | | | |
| Access | 0.009*** | [0.004] | 0.9% | 0.008** | [0.003] | 0.8% |
| food | 0.017*** | [0.004] | 1.7% | 0.017*** | [0.005] | 1.7% |
| **Controlling Factors** | | | | | | |
| northmpls | -0.472** | [0.182] | -37.6% | -0.544*** | [0.190] | -42.0% |
| open2010 | 0.481*** | [0.124] | 61.8% | 0.505*** | [0.123] | 65.7% |
| cclrt | -0.482*** | [0.168] | -38.2% | -0.522*** | [0.171] | -40.7% |
| opendate | - | - | - | (exposure) | | |
| **Constant** | 1.426*** | [0.446] | N/A | 1.582*** | [0.408] | N/A |
| | | | | | | |
| Dispersion Factor (α) | - | - | - | 0.161*** | [0.021] | N/A |
| No. of observations | 116 | | | 116 | | |
| Joint significance | 80.14 (F-statistics), p = 0.0000 | | | 230.27 (LR chi-square), p = 0.0000 | | |
| R-squares | 0.847 (adjusted) | | | 0.863 (Cox-Snell Pseudo) | | |

Note: * p<0.1; ** p<0.05; ***p<0.01